\documentclass[prc,twocolumn,superscriptaddress,superscriptfootnote]{revtex4}
\usepackage{amssymb}

\usepackage{amsmath,amsfonts,amssymb,epsfig,color}

\begin{document}
\title{New Description of the Doublet Bands in Doubly Odd Nuclei}
\author{H. G. Ganev}
\affiliation{Institute of Nuclear Research and Nuclear Energy,
Bulgarian Academy of Sciences, \\ Sofia 1784, Bulgaria}
\author{A. I. Georgieva}
\affiliation{Institute of Nuclear Research and Nuclear Energy,
Bulgarian Academy of Sciences, \\ Sofia 1784, Bulgaria}
\author{S. Brant}
\affiliation{Department of Physics, Faculty of Science, University
of Zagreb, 10000 Zagreb, Croatia}
\author{A. Ventura}
\affiliation{Ente per le Nuove tecnologie, l'Energia e l'Ambiente,\\
I-40129 Bologna and Istituto Nazionale di Fisica Nucleare, \\
Sezione di Bologna, Italy}

\setcounter{MaxMatrixCols}{10}

\begin{abstract}
The experimentally observed $\Delta I = 1$ doublet bands in some
odd-odd nuclei are analyzed within the orthosymplectic extension of
the Interacting Vector Boson Model (IVBM). A new, purely collective
interpretation of these bands is given on the basis of the obtained
boson-fermion dynamical symmetry of the model. It is illustrated by
its  application to three odd-odd nuclei from the $A\sim 130$
region, namely $^{126}Pr$, $^{134}Pr$ and $^{132}La$. The
theoretical predictions for the energy levels of the doublet bands
as well as $E2$ and $M1$ transition probabilities between the states
of the yrast band in the last two nuclei are compared with
experiment and the results of other theoretical approaches. The
obtained results reveal the applicability of the orthosymplectic
extension of the IVBM.
\end{abstract}
\maketitle PACS {21.10.Re,23.20.Lv,21.60.Fw,27.60.+j}

\section{Introduction}

In recent years, extensive experimental evidence for the existence
of distinct band structures in odd-odd nuclei has been obtained. It
has created an opportunity for testing the predictions of different
theoretical models on the level properties of these nuclei. One such
study involves the observation of doublet $\Delta I = 1$ bands in
odd-odd $N=75$ and $N=73$ isotones in the $A\sim 130$ region. A
large number of experimental data \cite{a1}-\cite{a8} have been
accumulated in this mass  region, showing that the yrast and yrare
states with the $\pi h_{11/2} \otimes \nu h_{11/2}$ configuration
form $\Delta I = 1$ doublet bands which are nearly degenerate in
energy. They are built on the single particle states of a valence
neutron and a valence proton in the same unique-parity orbital
$0h_{11/2}$. Pairs of bands have been found also in the $A \sim 105$
and $A \sim 190$ mass regions. Initially, these $\Delta I = 1$
doublet bands had been interpreted as a manifestation of
``chirality'' in the sense of the angular momentum coupling
\cite{FM}. Several theoretical models have been applied in a number
of articles, like the tilted axis cranking (TAC) model
\cite{a8},\cite{tac1}-\cite{tac4}, the core-quasiparticle coupling
model \cite{cqpcm2}, the particle-rotor model (PRM)
\cite{prm1}-\cite{prm3}, two quasiparticle + triaxial rotor model
(TQPTR) \cite{tqptrm} , core-particle-hole coupling model (CPHCM)
\cite{a6}. All these models have one assumption in common, they
suppose a rigid triaxial core and hence support the interpretation
of the doublet bands of chiral structure. On the contrary, all
odd-odd nuclei in which twin bands have been observed have a
different characteristics in common, they are in regions  where
even-even nuclei are $\gamma$-soft, i.e., effectively triaxial but
not rigid. Their potential energy surface is rather flat in the
$\gamma$-direction and the couplings with other core structures, not
only the ground state band, are significant. It is evident that
odd-odd nuclei in these mass regions do not satisfy all the
requirements for the existence of chirality, but they can approach
some of them, or at least retain some fingerprints of chirality.

Many of the recent experiments and theoretical analysis do not
support completely the chiral interpretation \cite{pr134}-\cite{e5}.
In particular, in an ideal situation, i.e. perfectly orthogonal
angular momentum vectors and stable triaxial nuclear shape, a
perfect degeneracy between the identical spin states should be
observed. In fact, the attaintment of degeneracy is one of the key
characteristics of chirality. This feature has not been observed in
any of the chiral structures identified to date. Moreover, states
with different quantum numbers in two nonchiral bands can also show
an accidental degeneracy. Thus, one of the important test of
chirality is that the degenerate states in the two bands should also
have similar physical properties, such as moment of inertia,
quasiparticle alignments, transition quadrupole moments, and the
related $B(E2)$ values for intraband $E2$ transitions. Some
experimental studies have shown that the two bands have different
shapes due to the different kinematical moments of inertia, which
suggest a shape coexistence (triaxial and axial shapes). This is an
interesting observation since the quantal nature of chirality
automatically demands that a chiral partner band should have
identical properties to the yrast triaxial rotational band.
Similarly, it was also found  that the experimental data for the
behavior of other observables (equal $E2$ transitions, staggering
behavior of the $M1$ values, the smoothness of the signature $S(I)$,
etc.) do not support such a chiral structure \cite{pr134}-\cite{e5}.
These results demand a deeper and more detailed discussion of our
understanding of the origin of doublet bands.

Within the framework of pair truncated shell model it was pointed
out that the band structure of the doublet bands can be explained by
the chopsticks-like motion of two angular momenta of the odd neutron
and the odd proton \cite{ptsm1}-\cite{ptsm3}.  It was found that the
level scheme of $\Delta I = 1$ doublet bands does not arise from the
chiral structure, but from different angular momentum configurations
of the unpaired neutron and unpaired proton in the $0h_{11/2}$
orbitals, weakly coupled with the collective excitations of the
even-even core. The same interpretation was given also in the
quadrupole coupling model \cite{qcm1},\cite{qcm2}.

An alternative interpretation has been based on the Interacting
boson fermion-fermion model (IBFFM) \cite{IBFFMa},\cite{IBFFMb},
where the energy degeneracy is obtained but a different nature is
attributed to the two bands. A detailed analysis of the wave
functions in IBFFM showed as well that the presence of
configurations with the angular momenta of the proton, neutron and
core in the chirality favorable, almost orthogonal geometry, is
substantial but far from being dominant. The large fluctuations of
the deformation parameters $\beta$ and $\gamma$ around the triaxial
equilibrium shape enhance the content of achiral configurations in
the wave functions. The $\beta-$distribution of the yrast band has
its maximum at larger deformations than that of the side band. At
higher angular momenta, this difference becomes very pronounced. In
addition, the fluctuations of $\beta$ in the side band become very
large with increasing spin. In both bands the fluctuations of
$\gamma$ increase with spin, being more pronounced in the side band
\cite{DC1}. The composition of the yrast band, in terms of
contributions from core states, shows that the yrast band is
basically built on the ground-state band of the even-even core. With
increasing spin the admixture of the $\gamma-$band of the core
becomes more pronounced. The side band wave functions contain large
components of the $\gamma-$band and with increasing spin, of
higher-lying collective structures of the core, which near the band
crossing become dominant. So, the conclusion of Refs.
\cite{e4},\cite{DC1} was that the existence of twin bands in
$^{134}Pr$ should be attributed to a weak dynamic (fluctuation
dominated) chirality combined with an intrinsic symmetry yet to be
revealed. The IBFFM was applied to the doublet bands in $^{134}Pr$
\cite{pr134},\cite{e4},\cite{DC1}.  The $B(E2)$ values of the
transitions depopulating the analog states are different from the
chiral predictions and  the $B(M1)$ staggering is not present
\cite{DC2}. The IBFFM was also applied for the description of the
yrast $\pi h_{11/2} \otimes \nu h_{11/2}$ band in $^{126}Pr$
\cite{pr126}.

The above variety of models and approaches dealing with the
description of the doublet bands in odd-odd nuclei motivated us to
consider their properties in the framework of the boson-fermion
extension of the symplectic IVBM \cite{GGG}.

In the present work we carry out an analysis of the doublet bands in
some doubly odd nuclei from the $A\sim 130$ region within the
orthosymplectic extension \cite{OSE} of the IVBM. The latter was
proposed in order to encompass the treatment of the odd-mass nuclei.
Further, the  new version of IVBM was applied for the description of
the ground and first excited positive and/or negative bands of
odd-odd nuclei \cite{OON}. The spectrum of the positive-parity
states in the odd-odd nuclei considered in this paper is based on
the odd proton and odd neutron (both particle-like in contrast to
usually considered proton particle-like and neutron hole-like nature
of the two odd particles) which occupy the same single particle
level $h_{11/2}$. The theoretical description of the doubly odd
nuclei under consideration is fully consistent and starts with the
calculation of theirs even-even and odd-even neighbors. We consider
the simplest physical picture in which two particles (or
quasiparticles) with intrinsic spins taking a single $j-$value are
coupled to an even-even core nucleus whose states belong to an
$Sp(12,R)$ irreducible representation. Thus, the bands of the
odd-mass and odd-odd nuclei arise as collective bands build on a
given even-even nucleus. So, within the framework of the
orthosymplectic extension of the model a purely collective structure
of the doublet bands is obtained.

The level structure of $^{126}Pr$, $^{134}Pr$ and $^{132}La$ is
analyzed in the framework of the orthosymplectic extension of the
IVBM \cite{OSE}. Thus to describe the structure of odd-odd nuclei,
first a description of the appropriate even-even cores should be
obtained.

\section{The even-even core nuclei}

The algebraic structure of the IVBM is realized in terms of creation
and annihilation operators $u_{m}^{+}(\alpha )$, $u_{m}(\alpha )$
($m=0,\pm 1$). The bilinear products of the creation and
annihilation operators of the two vector bosons generate the boson
representations of the non-compact symplectic group $ Sp(12,R)$
\cite{IVBM}:
\begin{eqnarray}
F_{M}^{L}(\alpha ,\beta ) &=& {\sum }_{k,m}C_{1k1m}^{LM}u_{k}^{+}(
\alpha )u_{m}^{+}(\beta ),  \nonumber \\
G_{M}^{L}(\alpha ,\beta ) &=&{\sum }_{k,m}C_{1k1m}^{LM}u_{k}(\alpha
)u_{m}(\beta ),  \label{pairgen}
\end{eqnarray}
\begin{equation}
A_{M}^{L}(\alpha, \beta )={\sum }_{k,m}C_{1k1m}^{LM}u_{k}^{+}(\alpha
)u_{m}(\beta ),  \label{numgen}
\end{equation}
where $C_{1k1m}^{LM}$, which are the usual Clebsch-Gordan
coefficients for $L=0,1,2$ and $M=-L,-L+1,...L$, define the
transformation properties of (\ref{pairgen}) and (\ref{numgen})
under rotations. The commutation relations between the pair creation
and annihilation operators \ (\ref{pairgen}) and the number
preserving operators (\ref{numgen}) are given in \cite{IVBM}.

Being a noncompact group, the unitary representations of $Sp(12,R)$
are of infinite dimension, which makes it impossible to diagonalize
the most general Hamiltonian. When reduced to the group $U^{B}(6)$,
each irrep of the group $Sp^{B}(12,R)$ decomposes into irreps of the
subgroup characterized by the partitions \cite{GGG},\cite{Q1}:
\begin{equation*}
\lbrack N,{0}^5\rbrack_{6}\equiv \lbrack N]_{6},\qquad
\end{equation*}%
where $N=0,2,4,\ldots$ (even irrep) or $N=1,3,5,\ldots$ (odd irrep).
The subspaces $[N]_{6}$ are  finite dimensional, which simplifies
the problem of diagonalization. Therefore the complete spectrum of
the system can be calculated through the diagonalization of the
Hamiltonian in the subspaces of all the unitary irreducible
representations (UIR) of $U(6)$, belonging to a given UIR of
$Sp(12,R)$, which further clarifies its role of a group of dynamical
symmetry.

The Hamiltonian, corresponding to the unitary limit of IVBM
\cite{GGG}
\begin{equation}
Sp(12,R)\supset U(6)\supset U(3)\otimes U(2)\supset O(3)\otimes
(U(1)\otimes U(1)),  \label{uchain}
\end{equation}%
expressed in terms of the first and second order invariant operators
of the different subgroups in the chain (\ref{uchain}) is
\cite{GGG}:
\begin{equation}
H=aN+bN^{2}+\alpha _{3}T^{2}+\beta _{3}L^{2}+\alpha _{1}T_{0}^{2}.
\label{Hrot}
\end{equation}
$H$ (\ref {Hrot}) is obviously diagonal in the basis
\begin{equation}
\mid [N]_{6};(\lambda,\mu);KLM;T_{0}\rangle \equiv \ \mid
(N,T);KLM;T_{0}\rangle, \label{bast}
\end{equation}
labeled by the quantum numbers of the subgroups of the chain
(\ref{uchain}). Its eigenvalues are the energies of the basis states
of the boson representations of $ Sp(12,R)$:
\begin{eqnarray}
E((N,T),L,T_{0}) &=& aN + bN^{2} + \alpha_{3}T(T+1)  \nonumber \\
&+& \beta_{3}L(L+1)+\alpha_{1}T_{0}^{2}.  \label{Erot}
\end{eqnarray}

The construction of the symplectic basis for the even IR of
$Sp(12,R)$ is given in detail in \cite{GGG}. The $Sp(12,R)$
classification scheme for the $ SU(3)$ boson representations for
even value of the number of bosons $N$  is shown on Table I in Ref.
\cite{GGG} (see also Table \ref{BasTab}).

\begin{figure}[t]
\centerline{\epsfxsize=3.5in\epsfbox{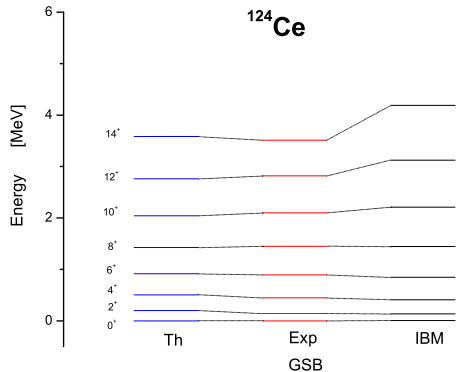}} \caption{(Color online)
Comparison of the theoretical and experimental energies for the
ground band of $^{124}$Ce.} \label{ce124}
\end{figure}

\begin{figure}[t]
\centerline{\epsfxsize=3.5in\epsfbox{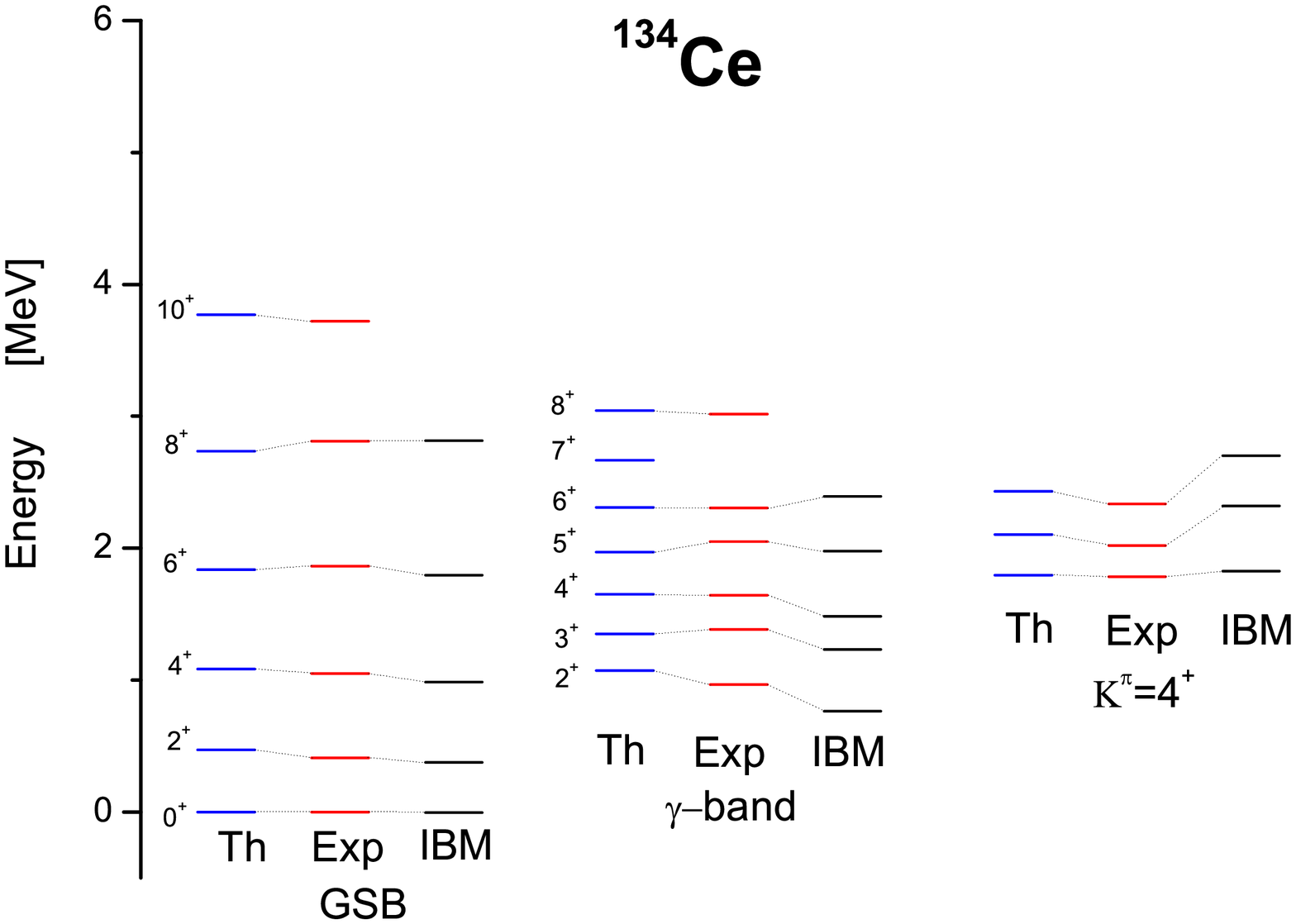}} \caption{(Color online)
Comparison of the theoretical and experimental energies for the
ground and first excited bands of $^{134}$Ce.} \label{ce134}
\end{figure}

\begin{figure}[t]
\centerline{\epsfxsize=3.5in\epsfbox{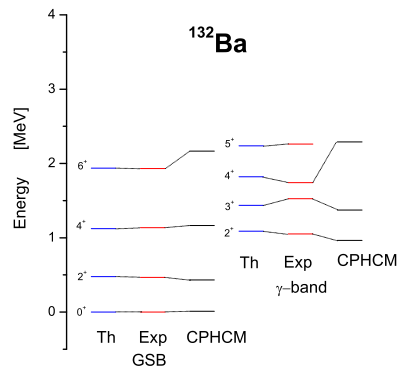}} \caption{(Color online)
Comparison of the theoretical and experimental energies for the
ground and $\gamma$ bands of $^{132}$Ba.} \label{ba132}
\end{figure}

The most important application of the $U^{B}(6)\subset $
$Sp^{B}(12,R)$ limit of the theory is the possibility it affords for
describing both even and odd parity bands up to very high angular
momentum \cite{GGG}. In order to do this we first have to identify
the experimentally observed bands with the sequences of basis states
of the even $Sp(12,R)$ irrep (Table \ref{BasTab}). As we deal with
the symplectic extension we are able to consider all even
eigenvalues of the number of vector bosons $N$ with the
corresponding set of $T-$spins, which uniquely define the
$SU^{B}(3)$ irreps $(\lambda, \mu)$. The multiplicity index $K$
appearing in the final reduction to the $SO(3)$ is related to the
projection of $L$ on the body fixed frame and is used with the
parity ($\pi $)\ to label the different bands ($K^{\pi } $) in the
energy spectra of the nuclei. For the even-even nuclei we have
defined the parity of the states as $\pi_{core} =(-1)^{T}$
\cite{GGG}. This allowed us to describe both positive and negative
bands.

Further, we use the algebraic concept of \textquotedblleft
yrast\textquotedblright\ states, introduced in \cite{GGG}. According
to this concept we consider as yrast states the states with given
$L$, that minimize the energy (\ref{Erot}) with respect to the
number of vector bosons $N$ that build them. Thus the states of the
ground state band (GSB) were identified with the $SU(3)$ multiplets
$(0,\mu )$ \cite{GGG}. In terms of $(N,T)$ this choice corresponds
to $(N=2\mu ,T=0)$ and the sequence of states with different numbers
of bosons $N=0,4,8,\ldots $ and $T=0$, $T_{0}=0$. Hence the minimum
values of the energies (\ref{Erot}) are obtained at $N=2L$.

The presented mapping of the experimental states onto the $SU(3)$
basis states, using the algebraic notion of yrast states, is a
particular case of the so called "stretched" states \cite{str}. The
latter are defined as the states with ($\lambda_{0}+2k,\mu_{0}$) or
($\lambda_{0},\mu_{0}+ k$), where $N_{i}=\lambda_{0}+2\mu_{0}$ and
$k=0,1,2,3, \ldots$.

It was established \cite{GGD} that the correct placement of the
bands in the spectrum strongly depends on their bandheads'
configuration, and in particular, on the minimal or initial number
of bosons, $N = N_{i}$, from which they are built. The latter
determines the starting position of each excited band.

Thus, for the description of the different excited bands, we first
determine the $N_{i}$ of the band head structure and develop the
corresponding excited band over the stretched $SU(3)$ multiplets.
This corresponds to the sequence of basis states with
$N=N_{i},N_{i}+4,N_{i}+8,\ldots$ ($\Delta N=4$). The values of $T$
for the first type of stretched states ($\lambda-$changing) are
changing by step $\Delta T=2$, whereas for the second type
($\mu-$changing) $-T$ is fixed so that in both cases the parity is
preserved even or odd, respectively. For all presented even-even
nuclei, the states of the described excited bands are associated
with the stretched states of the first type ($\lambda-$ changing).

To describe the structure of odd-mass and odd-odd nuclei, first a
description of the appropriate even-even cores should be obtained.
We determine the values of the five phenomenological model
parameters $a, b, \alpha_{3}, \beta_{3}, \alpha_{1}$ by fitting the
energies of the ground and $\gamma-$ bands of the even-even nuclei
to the experimental data \cite{exp}, using a $\chi^{2}$ procedure.

Numerous IBM studies of even-even nuclei in the $A\sim 130$ mass
region have shown that these nuclei are well described by the $O(6)$
symmetry of the IBM, that in the classical limit corresponds to the
Wilets-Jean model of a $\gamma-$unstable rotor \cite{WJR}, and that
the accepted interpretation is that they are $\gamma-$soft. The core
nucleus $^{124}Ce$ follows the systematic trend of the Ce isotopes.
The heavier isotopes are $\gamma-$soft ($O(6)$-like in the IBM
terminology), and the lighter ones are considerably deformed, but
they never reach the rigid rotor structure which corresponds to the
$SU(3)$ limit of the IBM. The transition between these two
structures occurs for $^{126}Ce$, and is reflected in the dynamics
of bands in the neighboring odd-even and odd-odd nuclei. In contrast
to the $O(6)$-like spectra observed in the odd-odd isotopes
$^{130,132}Pr$ \cite{Pr130132}, the structure of $^{126}Pr$ reflects
the transitional $SU(3)$-$O(6)$ nature of the core nucleus
$^{124}Ce$.

Here, we must point out that only in the considered dynamical
symmetry (\ref{uchain}) of the IVBM, due to the employed ``algebraic
yrast'' condition $N=2L$ and the reduction rules connecting the
values of the number of bosons $N$ with their angular momentum $L$
the energies of the collective states of ground state band
\cite{GGG} for example, can be written as:
\begin{equation}
E_g(L) = (2a-4b)L + (4b+ \beta_{3})L(L+1) \ ,  \label{Eg}
\end{equation}
where obviously  the rotational $L(L+1)$ and vibrational $L$
collective modes are mixed  and the type of collectivity depends on
the ratio of the coefficients in front of these two terms. In
analogy it could be shown that the two collective modes are mixed in
the excited bands as well. Hence we can describe quite well in the
same group chain of the symplectic extension, the even-even cores
with various collective properties that need different dynamical
symmetries or their mixture in the IBM.

The theoretical predictions for the even-even core nuclei are
presented in Figures \ref{ce124}$-$\ref{ba132}. For comparison, the
predictions of IBM and CPHCM are also shown. The IBM results for
$^{124}Ce$ and $^{134}Ce$ are extracted from Refs.
\cite{pr126},\cite{pr134} and those of CPHCM for $^{132}Ba$ from
\cite{a6}, respectively. From the figures one can see that the
calculated energy levels  of both ground state and $\gamma$ bands
agree rather well up to high angular momenta with the observed data.
Except for the GSB of $^{134}Ce$, for which the IVBM and IBM results
are almost identical, the IVBM predictions reproduce better the band
structures compared to CPHCM and IBM.

\section{Fermion degrees of freedom}

In order to incorporate the intrinsic spin degrees of freedom into
the symplectic IVBM, we extend the dynamical algebra of $Sp(12,R)$
to the orthosymplectic algebra of $OSp(2\Omega/12,R)$ \cite{OSE}.
For this purpose we introduce a particle (quasiparticle) with spin
$j$ and consider a simple core plus particle picture. Thus, in
addition to the boson collective degrees of freedom (described by
dynamical symmetry group $Sp(12,R)$) we introduce creation and
annihilation operators $a_{m}^{\dag}$ and $a_{m}$ ($m=-j,\ldots,j$),
which satisfy the anticommutation relations
\begin{eqnarray}
\{a_{m}^{\dag},a_{m'}^{\dag}\} &=&\{a_{m},a_{m'}\}=0,  \nonumber \\
\{a_{m},a_{m'}^{\dag}\} &=&\delta _{mm'}. \label{anticom}
\end{eqnarray}%
All bilinear combinations of \ $a_{m}^{+}$ and $a_{m'}$, namely
\begin{eqnarray}
f_{mm'} &=&a_{m}^{\dag}a_{m'}^{\dag}, \ \ m\neq m' \nonumber \\
g_{mm'} &=&a_{m}a_{m'}, \ \ m\neq m'; \label{O2N gen} \\
C_{mm'} &=&(a_{m}^{\dag}a_{m'}-a_{m'}a_{m}^{\dag})/2  \label{UN gen}
\end{eqnarray}%
generate the (Lie) fermion pair algebra of $SO^{F}(2\Omega)$. Their
commutation relations are given in \cite{OSE}. The number preserving
operators (\ref{UN gen}) generate maximal compact subalgebra
$U^{F}(\Omega)$ of $SO^{F}(2\Omega)$. The upper script $B$ or $F$
denotes the boson or fermion degrees of freedom, respectively.

\subsection{Fermion dynamical symmetries}

As can be seen from (\ref{UN gen}), the full number conserving
symmetry of a fermion of spin $j$ is $U^{F}(2j+1)$. In general, the
full dynamical algebra build from all bilinear combinations
(\ref{O2N gen}),(\ref{UN gen}) of creation and annihilation fermion
operators is the $SO(2\Omega)$ algebra (for a multilevel case
$\Omega=\sum_{j}(2j+1)$). One can further construct a certain
fermion dynamical symmetry, i.e. the group-subgroup chain:
\begin{equation}
SO(2\Omega) \supset  G ' \supset  G '' \supset \ldots .\label{FDS}
\end{equation}
In particular for one particle occupying a single level $j$ we are
interested in the following dynamical symmetry:
\begin{equation}
SO^{F}(2\Omega) \supset  Sp(2j+1) \supset SU^{F}(2),\label{opFDS}
\end{equation}
where $Sp(2j+1)$ is the compact symplectic group. The dynamical
symmetry (\ref{opFDS}) remains valid also for the case of two
particles occupying the same level $j$. In this case, the allowed
values of the quantum number $I$ of $SU(2)$ in (\ref{opFDS})
according to reduction rules are $I=0,2,\ldots,2j-1$ \cite{pair}. If
the two particles occupy different levels $j_{1}$ and $j_{2}$ of the
same or different major shell(s), one can consider the chain
\begin{widetext}
\begin{equation}
SO(2\Omega )\supset U(\Omega )%
\begin{tabular}{llllllll}
$\nearrow $ & $U(\Omega _{1})$ & $\supset $ & $Sp(2j_{1}+1)$ & $\supset $ & $%
SU_{I_{1}}(2)$ & $\searrow $ &  \\
&  &  &  &  &  &  & $SU^{F}(2)$ \\
$\searrow $ & $U(\Omega _{2})$ & $\supset $ & $Sp(2j_{2}+1)$ & $\supset $ & $%
SU_{I_{2}}(2)$ & $\nearrow $ &
\end{tabular}
\label{tpFDS}
\end{equation}
\end{widetext}
where $\Omega=\Omega_{1} + \Omega_{2}$. We want to point out that
although the final group $SU^{F}(2)$ that appears in the chain
(\ref{tpFDS}) is the same as in (\ref{opFDS}), its content is
different. Here the values of the common fermion angular momentum
$I$ are determined by the vector sum of the two individual spins
$I_{1}$ and $I_{2}$, respectively. Nevertheless, for simplicity
hereafter we will use just the reduction $SO(2\Omega) \supset
SU^{F}(2)$ (i.e. dropping all intermediate subgroups between
$SO(2\Omega)$ and $SU^{F}(2)$) and keep in mind the proper content
of the set of $I$ values for one and/or two particles cases,
respectively.

\subsection{Bose-Fermi symmetry}

Once the fermion dynamical symmetry is determined we proceed with
the construction of the Bose-Fermi symmetries. If a fermion is
coupled to a boson system having itself a dynamical symmetry (e.g.,
such as an IBM core), the full symmetry of the combined system is
$G^{B} \otimes G^{F}$. Bose-Fermi symmetries occur if at some point
the same group appears in both chains
\begin{equation}
G^{B} \otimes G^{F} \supset G^{BF},\label{BFS}
\end{equation}
i.e. the two subgroup chains merge into one. It should be noted that
(\ref{BFS}) is true only for the diagonal subgroup $G^{B} \otimes
G^{F}$, i.e. the one in which the two group elements multiplied
directly are parametrized by the same parameters. In this way the
Bose-Fermi symmetry not only constrains parameters by the choice of
particular subgroup chains in the boson and fermion sectors, but
also specifies the interaction between the two.

\section{Dynamical supersymmetry}

The standard approach to supersymmetry in nuclei (dynamical
supersymmetry) is to embed the Bose-Fermi subgroup chain of $G^{B}
\otimes G^{F}$ into a larger supergroup G, i.e. $G \supset G^{B}
\otimes G^{F}$. It is our intention in this paper to do that for
chains describing odd-odd nuclei.

Making use of the embedding $SU^{F}(2)\subset SO^{F}(2\Omega)$ and
considerations from the preceding section, we make orthosymplectic
(supersymmetric) extension of the IVBM which is defined through
the chain \cite{OSE}:
\begin{equation}
\begin{tabular}{lllll}
$OSp(2\Omega/12,R)$ & $\supset $ & $SO^{F}(2\Omega)$ & $\otimes $ & $Sp^{B}(12,R)$ \\
&  &  &  & $\ \ \ \ \ \ \Downarrow $ \\
&  & $\ \ \ \ \ \Downarrow $ & $\otimes $ & \ $\ U^{B}(6)$ \\
&  &  &  & $\ \ \ \ \ N$ \\
&  &  &  & $\ \ \ \ \ \ \Downarrow $ \\
&  & $SU^{F}(2)$ & $\otimes $ & $SU^{B}(3)\otimes U_{T}^{B}(2)$ \\
&  & $\ \ \ \ \ I$ &  & $(\lambda ,\mu )\Longleftrightarrow (N,T)~$ \\
&  & \multicolumn{1}{r}{$\searrow $} &  & $\ \ \ \ \ \ \Downarrow $ \\
&  &  & $\otimes $ & $SO^{B}(3)\otimes U(1)$ \\
&  &  &  & $~~~L\qquad \qquad T_{0}$ \\
&  &  & $\Downarrow $ &  \\
&  & $Spin^{BF}(3)$ & $\supset $ & $Spin^{BF}(2),$ \\
&  & $~~~~~J$ &  & $~~~~~J_{0}$%
\end{tabular}
\label{chain}
\end{equation}%
where below the different subgroups the quantum numbers
characterizing their irreducible representations are given.
$Spin^{BF}(n)$ ($n=2,3$) denotes the universal covering group of
$SO(n)$.

In the next section we present the application of the boson-fermion
extension of IVBM, developed for the description of the collective
bands of even-even \cite{GGG} and odd-mass \cite{OSE} nuclei, in
order to include in our considerations the positive parity states of
the yrast and side bands of odd-odd nuclei from $A\sim 130$ region,
build on $\pi h_{11/2} \otimes \nu h_{11/2}$ configuration.

\section{The energy spectra of odd-mass and odd-odd nuclei}

We can label the basis states according to the chain (\ref{chain})
as:
\begin{eqnarray}
&&|~[N]_{6};(\lambda,\mu);KL;I;JJ_{0};T_{0}~\rangle \equiv \nonumber \\
&& |~[N]_{6};(N,T);KL;I;JJ_{0};T_{0}~\rangle ,  \label{Basis}
\end{eqnarray}%
where $[N]_{6}-$the $U(6)$ labeling quantum number and
$(\lambda,\mu)-$the $ SU(3)$ quantum numbers characterize the core
excitations, $K$ is the multiplicity index in the reduction
$SU(3)\supset SO(3)$, $L$ is the core angular momentum, $I-$the
intrinsic spin of an odd particle (or the common spin of two fermion
particles for the case of odd-odd nuclei), $J,J_{0}$ are the total
(coupled boson-fermion) angular momentum and its third projection,
and $T$,$T_{0}$ are the $T-$spin and its third projection,
respectively. Since the $SO(2\Omega)$ label is irrelevant for our
application, we drop it in the states (\ref{Basis}).

The Hamiltonian of the combined boson-fermion system can be written
as linear combination of the Casimir operators of the different
subgroups in (\ref{chain}): \
\begin{eqnarray}
H &=& aN+bN^{2}+\alpha _{3}T^{2}+\beta _{3}^{\prime
}L^{2}+\alpha_{1}T_{0}^{2} \nonumber \\
&+& \eta I^{2}+\gamma ^{\prime }J^{2}+\zeta J_{0}^{2}
\label{Hamiltonian}
\end{eqnarray}
and it is obviously diagonal in the basis (\ref{Basis}) labeled by
the quantum numbers of their representations. Then the eigenvalues
of the Hamiltonian (\ref{Hamiltonian}), that yield the spectrum of
the odd-mass and odd-odd systems are:
\begin{eqnarray}
&&E(N;T,T_{0};L,I;J,J_{0})= \nonumber \\
&&aN+bN^{2} \nonumber \\
&&+\alpha _{3}T(T+1)+
\beta _{3}^{\prime }L(L+1)+\alpha _{1}T_{0}^{2} \nonumber \\
&&+\eta I(I+1)+\gamma ^{\prime }J(J+1)+\zeta
J_{0}^{2}.\label{Energy}
\end{eqnarray}
We note that only the last three terms of (\ref{Hamiltonian}) come
from the orthosymplectic extension. We choose parameters $\beta _{3}
^{\prime }=\frac{1}{2}\beta _{3}$ and $\gamma ^{\prime
}=\frac{1}{2}\gamma$ instead of $\beta _{3}$ and $\gamma$ in order
to obtain the Hamiltonian form of ref. \cite{GGG} (setting $\beta
_{3} =\gamma$), when for the case $I=0$ (hence $J=L$) we recover the
symplectic structure of the IVBM.

The infinite set of basis states classified according to the
reduction chain (\ref{chain}) are schematically shown in Table
\ref{BasTab}. The fourth and fifth columns show the $SO^{B}(3)$
content of the $SU^{B}(3)$ group, given by the standard Elliott's
reduction rules \cite{Elliott}, while in the next column are given
the possible values of the common angular momentum $J$, obtained by
coupling of the orbital momentum $L$ with the spin $I$. The latter
is vector coupling and hence all possible values of the total
angular momentum $J$ should be considered. For simplicity, only the
maximally aligned ($J=L+I$) and maximally antialigned ($J=L-I$)
states are illustrated in Table \ref{BasTab}.
\begin{center}
\begin{table}[t]
\caption{Classification scheme of basis states (\protect\ref{Basis})
according the decompositions given by the chain
(\protect\ref{chain}).}
\smallskip \centering{\footnotesize \renewcommand{\arraystretch}{1.25}
\begin{tabular}{l|l|l|l|l|l}
\hline\hline $N$ & $T$ & $(\lambda ,\mu )$ & $K$ & $L$ & \quad \quad
\quad \quad \quad \quad \quad $J=L\pm I$ \\ \hline\hline $0$ & $0$ &
$(0,0)$ & $0$ & $0$ & $I$ \\ \hline\hline $2$ & $1$ & $(2,0)$ & $0$
& $0,2$ & $I;\ 2\pm I$ \\ \cline{2-6} & $0$ & $(0,1)$ & $0$ & $1$ &
$1\pm I$ \\ \hline\hline & $2$ & $(4,0)$ & $0$ & $0,2,4$ & $I;\ 2\pm
I;4\pm I$ \\ \cline{2-6}
$4$ & $1$ & $(2,1)$ & $1$ & $1,2,3$ & $1\pm I;\ 2\pm I;\ 3\pm I$ \\
\cline{2-6} & $0$ & $(0,2)$ & $0$ & $0,2$ & $ I;\ 2\pm I$ \\
\hline\hline
& $3$ & $(6,0)$ & $0$ & $0,2,4,6$ & $ I;\ 2\pm I;\ 4\pm I;6\pm I$ \\
\cline{2-6} & $2$ & $(4,1)$ & $1$ & $1,2,3,4,5$ &
\begin{tabular}{l}
$1\pm I;\ 2\pm I;\ 3\pm I;$ \\
$4\pm I;\ 5\pm I$%
\end{tabular}
\\ \cline{2-6}
$6$ & $1$ & $(2,2)$ & $2$ & $2,3,4$ & $2\pm I;\ 3\pm I;\ 4\pm I$ \\
\cline{2-6} &  &  & $0$ & $0,2$ & $ I;\ 2\pm I$ \\ \cline{2-6} & $0$
& $(0,3)$ & $0$ & $1,3$ & $1\pm I;\ 3\pm I$ \\ \hline\hline & $4$ &
$(8,0)$ & $0$ & $0,2,4,6,8$ &
\begin{tabular}{l}
$ I;\ 2\pm I;\ 4\pm I;$ \\
$6\pm I;\ 8\pm I$%
\end{tabular}
\\ \cline{2-6}
& $3$ & $(6,1)$ & $1$ & $1,2,3,4,5,6,7$ &
\begin{tabular}{l}
$1\pm I;\ 2\pm I;\ 3\pm I;$ \\
$4\pm I;\ 5\pm I;6\pm I;\ $ \\
$7\pm I;8\pm I$%
\end{tabular}
\\ \cline{2-6}
& $2$ & $(4,2)$ & $2$ & $2,3,4,5,6$ &
\begin{tabular}{l}
$2\pm I;\ 3\pm I;\ 4\pm I;$ \\
$5\pm I;\ 6\pm I$%
\end{tabular}
\\ \cline{2-6}
$8$ &  &  & $0$ & $0,2,4$ & $ I;\ 2\pm I;\ 4\pm I$ \\ \cline{2-6}
& $1$ & $(2,3)$ & $2$ & $2,3,4,5$ & $2\pm I;\ 3\pm I;\ 4\pm I;5\pm I$ \\
\cline{2-6} &  &  & $0$ & $1,3$ & $1\pm I;\ 3\pm I$ \\ \cline{2-6} &
$0$ & $(0,4)$ & $0$ & $0,2,4$ & $ I;2\pm I;\ 4\pm I$ \\
\hline\hline
$\vdots $ & $\vdots $ & $\vdots $ & $\vdots $ & $\vdots $ & $\vdots $%
\end{tabular}
} \label{BasTab}
\end{table}
\end{center}

The basis states (\ref{Basis}) can be considered as a result of the
coupling of the orbital $\mid (N,T);KLM;T_{0}\rangle$ (\ref{bast})
and spin $\phi_{j\equiv I,m}$ wave functions. Then, if the parity of
the single particle is $\pi_{sp}$, the parity of the collective
states of the odd$-A$ nuclei will be $\pi=\pi_{core} \pi_{sp}$
\cite{OSE}. In analogy, one can write $\pi=\pi_{core} \pi_{sp}(1)
\pi_{sp}(2)$ for the case of odd-odd nuclei. Thus, the description
of the positive and/or negative parity bands requires only the
proper choice of the core band heads, on which the corresponding
single particle(s) is (are) coupled to, generating in this way the
different odd$-A$ (odd-odd) collective bands.

\begin{figure}[h!]
\centerline{\epsfxsize=3.5in\epsfbox{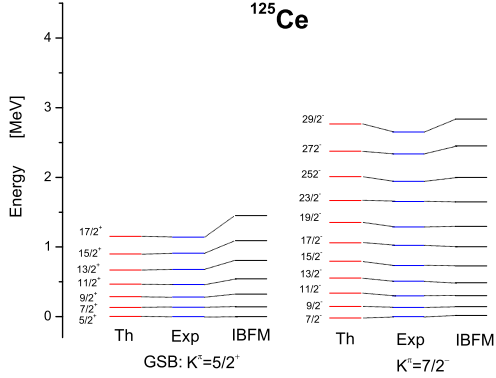}} \caption{(Color online)
Comparison of the theoretical and experimental energies for the
ground and first excited bands of $^{125}Ce$.} \label{ce125}
\end{figure}

\begin{figure}[h!]
\centerline{\epsfxsize=3.5in\epsfbox{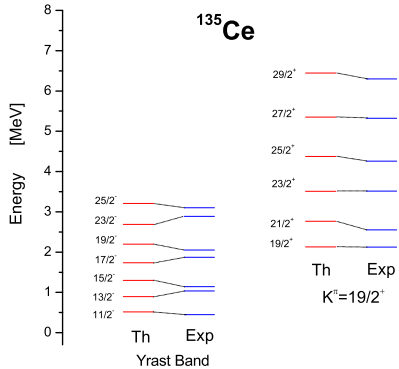}} \caption{(Color online)
The same as in the Fig. \protect\ref{ce125} but for $^{135}Ce$.}
\label{ce135}
\end{figure}

\begin{figure}[h!]
\centerline{\epsfxsize=3.5in\epsfbox{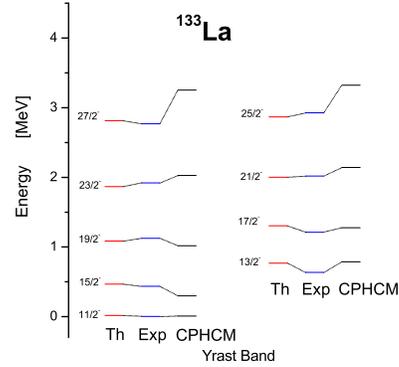}} \caption{(Color online)
Comparison of the theoretical and experimental energies for the
yrast band of $^{133}La$.} \label{la133}
\end{figure}

Further in the present considerations, the "yrast" conditions yield
relations between the number of bosons $N$ and the coupled angular
momentum $J$ that characterizes each collective state. For example,
the collective states of the GSB $K_{J} ^{\pi}=\frac{5}{2}^{+}$
($^{125}Ce$) of the odd-mass nuclei are identified with the $SU(3)$
multiplets $(0,\mu )$ which yield the sequence
$N=2(J-I)=0,2,4,\ldots$ for the corresponding values
$J=\frac{5}{2},\frac{7}{2},\frac{9}{2},...$. The T$-$spin for the
$SU(3)$ multiplets $(0,\mu )$ is $T=0$ and hence
$\pi_{core}=(-1)^{T}=(+)$. Here it is assumed that the single
particle has $j\equiv I=5/2$ and parity $\pi_{sp}=(+)$, so that the
common parity $\pi$ is also positive.

For the description of the different excited bands, we first
determine the $N_{i}$  of the band head structure and then we map
the states of the corresponding band onto the sequence of basis
states with $N=N_{i},N_{i}+2,N_{i}+4,\ldots$ ($\Delta N=2$) and
$T=even=fixed$ or $T=odd=fixed$, respectively. This choice
corresponds to the stretched states of the second type
($\mu-$changing).

The number of adjustable parameters needed for the complete
description of the collective spectra of both odd-A and odd-odd
nuclei is three, namely $\gamma$, $\zeta $ and $\eta$. The first two
are evaluated by a fit to the experimental data \cite{exp} of the
GSB of the corresponding odd-A neighbor, while the last one is
introduced in the final step of the fitting procedure for the
odd-odd nucleus, respectively. For the $A \sim 130$ region where the
doublet bands are built on $\pi h_{11/2} \otimes \nu h_{11/2}$
configuration, the two fermions occupy the same single particle
level $j_{1}=j_{2}=j=11/2$ with negative parity ($\pi_{sp}=-$) and
the fermion reduction chain (\ref{opFDS}) can be used.

The odd-A neighboring nuclei $^{125}Ce$ and $^{135}Ce$ can be
considered as a neutron coupled to the even-even cores $^{124}Ce$
and $^{134}Ce$, while the $^{133}La -$as a proton coupled to the
$^{132}Ba$, respectively. The low-lying positive parity states of
the GSB in odd-A neighbors are based on positive parity proton and
positive parity neutron configurations
($s_{\frac{1}{2}},d_{\frac{3}{2}},d_{\frac{5}{2}},g_{\frac{7}{2}}$),
whereas those of negative parity$-$ on $h_{11/2}$. In our
considerations we take into account only the first available single
particle orbit $j_{1}$ (generating the groups $SO(2\Omega_{1})$
and/or $U(\Omega_{1})$ with $\Omega_{1}=(2j_{1} +1)$).

The comparison between the experimental spectra for the GSB and
first excited band using the values of the model parameters given in
Table \ref{TablePar} for the nuclei $^{125}Ce$, $^{135}Ce$ and
$^{133}La$ is illustrated in Figures \ref{ce125}$-$\ref{la133}. One
can see from the figures that the calculated energy levels agree
rather well in general with the experimental data up to very high
angular momenta. For comparison, in Figures \ref{ce125} and
\ref{la133} the IBFM and CPHCM results for $^{125}Ce$ and $^{133}La$
are also shown. They are extracted from Refs. \cite{pr126} and
\cite{a6}, respectively.

For the calculation of the odd-odd nuclei spectra a second particle
should be coupled to the core. In our calculations a consistent
procedure is employed which includes the analysis of the even-even
and odd-even neighbors of the nucleus under consideration. Thus, as
a first step an odd particle was coupled to the boson core in order
to obtain the spectra of the odd-mass neighbors $^{125}Ce$,
$^{135}Ce$ and $^{133}La$. As a second step, we consider an addition
of a second particle to the boson-fermion system.

In our application, the most important point is the identification
of the experimentally observed states with a certain subset of basis
states from (ortho)symplectic extension of the model. Here we
consider a more general mapping when the states of the GSB of the
odd-odd nuclei are associated with a sequence of $SU(3)$ multiplets
$(0,\mu)$ but the band starts with the multiplet $(0,\mu_{0})$
instead of $(0,0)$. Thus, to the states of the yrast band with
$J=I,I+1,I+2,\ldots$ of the odd-odd nuclei we put into
correspondence the $SU(3)$ multiplets $(0,\mu)$ ($\mu-$changing) of
the basis states (\ref{Basis}) which in terms of $(N,T)$ correspond
to $(N=2\mu ,T=0)$ and the sequence of states with different numbers
of bosons $N=N_{0},N_{0}+2,N_{0}+4,\ldots$ ($\Delta N=2$). The
chosen set of $SU(3)$ multiplets $(0,\mu)$ means that the GSB of the
odd-odd as well as that of odd-mass nuclei is build on the GSB of
the even-even core nucleus. We recall that in contrast to the IBM,
the symplectic core structure (described by \emph{different} $SU(3)$
multiplets $(0,\mu)$) within the IVBM is active allowing the change
of the number of bosons. The "yrast" condition which results from
this mapping of the band's states over the stretched states
$(\lambda_{0}=0,\mu_{0}+k)$ yields $N=2\mu_{0}+2L$ (or $k=L$). In
particular, when the band head structure is determined by $N_{0}=0$
bosons, the yrast condition reduces to $N=2L$ (or $\mu=L$)
\cite{GGG},\cite{OSE}. In order to visualize the correspondence
under consideration, we illustrate the selected subset of basis
states in Table \ref{sub}. Hence one obtains the observed ground
state of the yrast band with $K^{\pi}_{J}=8^{+}$ for
$^{126}Pr$,$^{134}Pr$ and $^{132}La$ nuclei simply attributing to it
only the angular momentum $I=8$ from the vector coupling of the
proton $I_{p}=\frac{11}{2}$ and neutron $I_{n}=\frac{11}{2}$
momenta.

\begin{center}
\begin{table}[h]
\caption{The subset of basis states (\protect\ref{Basis}) associated
with the states of the GSB of odd-odd nuclei, based on $\pi h_{11/2}
\otimes \nu h_{11/2}$ configuration.}
\begin{tabular}{||l||l|l|l|l|l||}
\hline\hline $N$ & $N_{0}$ & $N_{0}+2$ & $N_{0}+4$ & $N_{0}+6$ &
$\ldots $ \\ \hline
$(\lambda ,\mu )$ & $(0,\mu _{0})$ & $(0,\mu _{0}+1)$ & $(0,\mu _{0}+2)$ & $%
(0,\mu _{0}+3)$ & $\ldots $ \\ \hline $L$ & $0$ & $1$ & $2$ & $3$ &
$\ldots $ \\ \hline $J$ & $I$ & $I+1$ & $I+2$ & $I+3$ & $\ldots $ \\
\hline\hline
\end{tabular}
\label{sub}
\end{table}
\end{center}

For the description of the side (yrare) band build also on the $\pi
h_{11/2} \otimes \nu h_{11/2}$ configuration which can be considered
as an excited band, we first determine the collective structure of
the band head $N_{i}=\lambda_{0}+2\mu_{0}$ and then map the states
of this band onto the sequences of basis states with
$N=N_{i},N_{i}+2,N_{i}+4,\ldots$ ($\Delta N=2$) and $T=even=fixed$.
This choice corresponds to the stretched states of the second type
($\mu-$changing). The $SU(3)$ multiplets $(\lambda \neq 0,\mu)$
attributed to the side band suggest similar collective structure for
this band compared to that of its "doublet partner". Similar
interpretation of the two bands takes place in the IBFFM, where the
yrast band is basically build on the GSB (described within \emph{a
single} $SU(3)$ multiplet $(\lambda,\mu)$) of the even-even core,
while the structure of the side band is that of odd proton and odd
neutron coupled to the $\gamma-$band of the core and in the high
spin region contains sizeable components of the higher-lying core
structures .

\begin{figure}[h!]
\centerline{\epsfxsize=3.5in\epsfbox{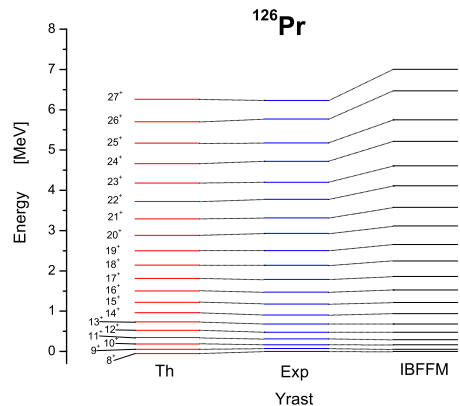}} \caption{(Color online)
Comparison of the theoretical and experimental energies for the
yrast band of $^{126}Pr$.} \label{pr126}
\end{figure}

\begin{figure}[h!]
\centerline{\epsfxsize=3.5in\epsfbox{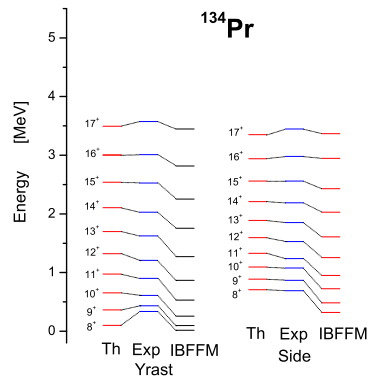}} \caption{(Color online)
Comparison of the theoretical and experimental energies for the
yrast and side bands of $^{134}Pr$.} \label{pr134}
\end{figure}

\begin{figure}[h!]
\centerline{\epsfxsize=3.5in\epsfbox{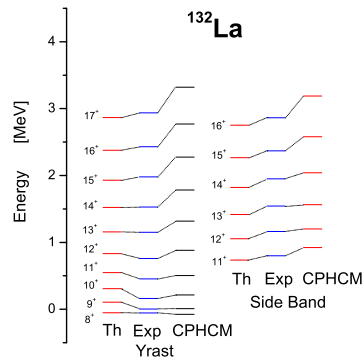}} \caption{(Color online)
The same as Fig. \ref{pr134}, but for $^{132}La$.} \label{la132}
\end{figure}

The theoretical predictions for the yrast and side bands based on
$\pi h_{11/2} \otimes \nu h_{11/2}$ configuration for the three
odd-odd nuclei $^{126}Pr$, $^{134}Pr$ and $^{132}La$ from $A \sim
130$ region are presented in Figures \ref{pr126},\ref{pr134} and
\ref{la132}, respectively. For comparison, the IBFFM
(refs.\cite{pr126},\cite{DC1}) and CPHCM (ref.\cite{a6}) results are
also shown. In Table \ref{TablePar}, the values of $N_{i}$, $T$,
$T_{0}$, $J$ and $\chi^{2}$ for each band under consideration are
also given. From the figures one can see the good overall agreement
between the theory and experiment which reveals the applicability of
the boson-fermion extension of the model.

\begin{table}[tb]
\caption{{Values of the model parameters.}}
\centering
\begin{tabular}{||l|l|l|l|l|l|l|l||}
\hline\hline $Nucl.$ & $bands$ & $N_{i}$ & $T$ & $T_{0}$ & $J$ &
$\chi ^{2}$ & $parameters $ \\ \hline\hline
$^{126}\Pr $ & $%
\begin{tabular}{l}
$Yrast:\allowbreak $ \\
$K^{\pi }=8^{+}$%
\end{tabular}%
$ & $24$ & $0$ & $0$ & $L+I$ & 0.0017 & $%
\begin{tabular}{l}
$a=$0.02855 \\
$b=-$0.00120 \\
$\alpha _{3}=$0.00680%
\end{tabular}%
$ \\ \hline $I=8$ &  &  &  &  &  &  &
\begin{tabular}{l}
$\beta _{3}=$0.01774 \\
$\alpha _{1}=$0.01387%
\end{tabular}
\\ \hline
&  &  &  &  &  &  &
\begin{tabular}{l}
$\eta =-$0.00906 \\
$\gamma =$0.01691 \\
$\zeta =-$0.01132%
\end{tabular}
\\ \hline\hline
$^{132}La$ & $%
\begin{tabular}{l}
$Yrast:\allowbreak $ \\
$K^{\pi }=8^{+}$%
\end{tabular}%
$ & $44$ & $0$ & $0$ & $L-I$ & 0.0034 &
\begin{tabular}{l}
$a=$0.07449 \\
$b=$0.00690 \\
$\alpha _{3}=$0.05709%
\end{tabular}
\\ \hline
$I=8$ & $%
\begin{tabular}{l}
$side:\allowbreak $ \\
$K^{\pi }=11^{+}$%
\end{tabular}%
$ & $50$ & $2$ & $0$ & $L-I$ & 0.0088 &
\begin{tabular}{l}
$\beta _{3}=$0.04847 \\
$\alpha _{1}=$0.06076%
\end{tabular}
\\ \hline
&  &  &  &  &  &  &
\begin{tabular}{l}
$\eta =$0.02360 \\
$\gamma =$0.04796 \\
$\zeta =$0.02960%
\end{tabular}
\\ \hline\hline
$^{134}\Pr $ & $%
\begin{tabular}{l}
$Yrast:\allowbreak $ \\
$K^{\pi }=8^{+}$%
\end{tabular}%
$ & $10$ & $0$ & $0$ & $L+I$ & 0.0046 &
\begin{tabular}{l}
$a=$0.08190 \\
$b=$0.00473 \\
$\alpha _{3}=$0.03637%
\end{tabular}
\\ \hline
$I=8$ & $\allowbreak
\begin{tabular}{l}
$side:$ \\
$K^{\pi }=8^{+}$%
\end{tabular}%
$ & $14$ & $4$ & $0$ & $L+I$ & 0.0020 &
\begin{tabular}{l}
$\beta _{3}=$0.03660 \\
$\alpha _{1}=$0.04424%
\end{tabular}
\\ \hline
&  &  &  &  &  &  &
\begin{tabular}{l}
$\eta =-$0.01876 \\
$\gamma =$0.03002 \\
$\zeta =$0.00061%
\end{tabular}
\\ \hline\hline
\end{tabular}
\label{TablePar}
\end{table}

To investigate the structure of the doublet bands in a certain
nucleus, it is crucial to determine the $B(E2)$ and $B(M1)$ values
which are very important for establishing the nature of these bands.
So, in the next section we consider the $E2$ and $M1$ transitions in
the framework of the orthosymplectic extension of the IVBM.

\section{Electromagnetic transitions}

A successful nuclear model must yield a good description not only of
the energy spectrum of the nucleus but also of its electromagnetic
properties. Calculation of the latter is a good test of the nuclear
model functions. The most important electromagnetic features which
manifest themselves in doublet bands are the $E2$ and $M1$
transitions. In this section we discuss the calculation of the $E2$
and $M1$ transition strengths between the states of the yrast band
of the odd-odd nuclei based on $\pi h_{11/2} \otimes \nu h_{11/2}$
configuration and compare the results with the available
experimental data.

For a mixed systems of bosons and fermions it is convenient to
expand the coupled basis states (\ref{Basis}) into  the direct
product of the boson and fermion states. The latter significantly
simplifies the application of the Wigner-Eckart theorem in the
practical calculations of the transition rates.

\subsection{E2 transitions}

As was mentioned, in the symplectic extension of the IVBM the
complete spectrum of the system is obtained in all the even
subspaces with fixed $N$- even of the UIR $[N]_{6}$ of $U(6)$,
belonging to a given even UIR of $Sp(12,R)$. The classification
scheme of the $SU(3)$ boson representations for even values of the
number of bosons $N$ was presented in Table \ref{BasTab}.

In the present paper, the states of the yrast band are identified
with the $SU(3)$ multiplets $(0,\mu )$. This yields the sequence
$N=N_{0},N_{0}+2,N_{0}+4,\ldots$ for the corresponding values
$J=I,I+1,I+2,...$ (see Table \ref{sub}). In terms of $(N,T)$ this
corresponds to $(N=2\mu ,T=0)$.

Using the tensorial properties of the $Sp(12,R)$ generators with
respect to (\ref{uchain}) it is easy to define the proper $E2$
transition operator between the states of the considered band as
\cite{TP}:
\begin{widetext}
\begin{equation}
T^{E2}=e\left[A_{(1,1)_{3}[0]_{2}\quad 00}^{ \lbrack 1-1]_{6}\quad
\quad 20} +\theta ([F\times F]_{(0,2)[0]_{2}\quad 00}^{\quad \lbrack
4]_{6}\quad \, \ 20}+[G\times G]_{(2,0)[0]_{2}\quad 00}^{\quad
\lbrack -4]_{6}\quad \,20}\right]. \label{te2}
\end{equation}%
\end{widetext}
The first part of (\ref{te2}) is a $SU(3)$ generator and actually
changes only the angular momentum with $\Delta L=2$.

The tensor product
\begin{widetext}
\begin{eqnarray} \label{FF}
\lbrack F\times F]_{(0,2)[0]_{2}\quad \, 00}^{\quad \lbrack
4]_{6}\quad \quad 20} &=&\sum C_{(2,0)[2]_{2} \, (2,0)[2]_{2}\quad
(0,2)[0]_{2}}^{\quad [2]_{6}\quad \quad \lbrack 2]_{6}\quad \quad
\lbrack
4]_{6}}C_{\,(2)_{3}\quad (2)_{3}\quad (2)_{3}}^{(2,0)\,\,(2,0)\quad(0,2)} \nonumber \\
&& \label{FF} \\
&&\times C_{20\ 20}^{20}C_{1 1\ 1 -1}^{10}\ F_{(2,0)[2]_{2}\ \
11}^{\quad \lbrack 2]_{6}\quad \ 20}F_{(2,0)[2]_{2}\ \ 1-1}^{\quad
\lbrack 2]_{6}\quad \ 20} \nonumber
\end{eqnarray}
\end{widetext}
of the operators (\ref{pairgen}) that are the pair raising
$Sp(12,R)$ generators changes the number of bosons by $\Delta N=4$
and $\Delta L=2$. It is obvious that this term in $T^{E2}$
(\ref{te2}) comes from the symplectic extension of the model. In
(\ref{te2}) $e$ is the effective boson charge.

The transition probabilities are by definition $SO(3)$ reduced
matrix elements of transition operators $T^{E2}$ (\ref{te2}) between
the $|i\rangle -$initial and $|f\rangle -$final collective states
(\ref{Basis})
\begin{equation}
B(E2;J_{i}\rightarrow J_{f})=\frac{1}{2J_{i}+1}\mid \langle \quad
f\parallel T^{E2}\parallel i\quad \rangle \mid ^{2}. \label{deftrpr}
\end{equation}
The basis states (\ref{Basis}) can be considered as a result of the
coupling of the orbital $\mid (N,T);KLM;T_{0}\rangle$ (\ref{bast})
and spin $\phi_{Im}$ wave functions. Since the spin $I$ ($I-fixed$)
is simply added to the orbital momentum $L$, the action of the
transition operator $T^{E2}$ concerns only the orbital part of the
basis functions (\ref{Basis}).

\begin{figure}[h!]
\centerline{\epsfxsize=3.5in\epsfbox{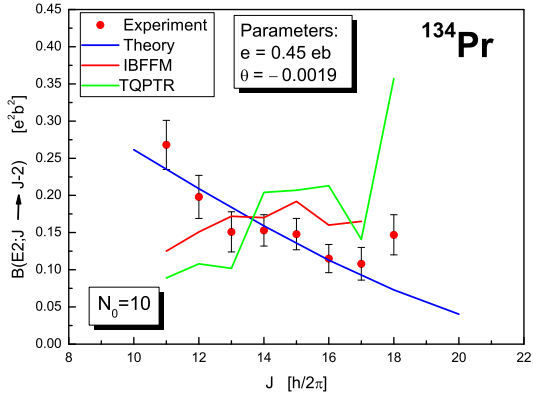}} \caption{(Color
online) Comparison of the theoretical and experimental values for
the $B(E2)$ transition probabilities for the $^{134}Pr$. The
theoretical predictions of the IBFFM and TQPTR are shown as well.}
\label{PrE2}
\end{figure}

In order to prove the correct predictions following from our
theoretical results we apply the theory to the two nuclei $^{134}Pr$
and $^{132}La$ for which there are available experimental data for
the transition probabilities between the states of the yrast bands.
The application actually consists of fitting the two parameters of
the transition operator $T^{E2}$ (\ref{te2}) to the experiment for
each of the considered bands. The $B(E2)$ strengths between the
positive parity states of the yrast band, as were attributed to the
$SU(3)$ symmetry-adapted basis states (\ref{Basis}) of the model,
are calculated. For these $SU(3)$ multiplets, the procedure for
their calculations actually coincides with that given in \cite{TP}
and modified for the case of odd-odd nuclei in \cite{ng}. The
theoretical predictions for the $^{134}Pr$ nucleus are compared with
the experimental data \cite{DC1} in Figure \ref{PrE2}. For
comparison, the IBFFM and TQPTR results (ref. \cite{DC1}) are also
shown. From the figure one can see the good overall reproduction of
the experimental values, which is obviously better than the IBFFM
and TQPTR ones.

\begin{figure}[h!]
\centerline{\epsfxsize=3.5in\epsfbox{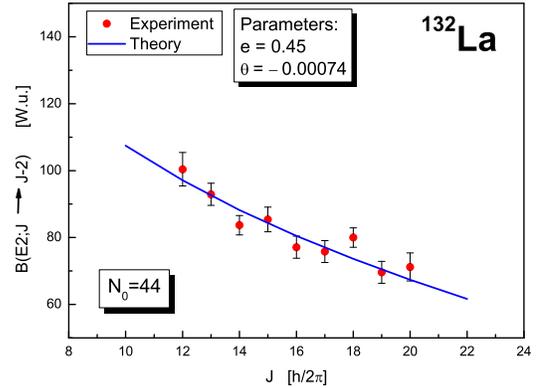}} \caption{(Color
online) Comparison of the theoretical and experimental values for
the $B(E2)$ transition probabilities for the $^{132}La$.}
\label{LaE2}
\end{figure}

In Figure \ref{LaE2} the theoretical predictions for the $^{132}La$
nucleus are compared with the experimental data \cite{e3}. One sees
that the experimental behavior of $B(E2)$ values of this nucleus is
also reproduced quite well.

\subsection{M1 transitions}

The structure of $M1$ transition operator between the states of the
yrast band can be defined in the following way:
\begin{widetext}
\begin{equation}
T^{M1 \quad(1)}_{\quad \quad M}=\sqrt{\frac{3}{4\pi}} \left[g
J_{M}^{(1)}+ g_{FG} (F_{(0,1)[0]_{2}\quad 00}^{\quad \lbrack
2]_{6}\quad \, \ 1M}+ G_{(1,0)[0]_{2}\quad 00}^{\quad \lbrack
-2]_{6}\quad \,1M}) \right]. \label{tm1}
\end{equation}
\end{widetext}
$J_{M}^{(1)}$ is the total boson-fermion angular
momentum, i.e. $J_{M}^{(1)}=L_{M}^{1}+ I_{M}^{(1)}$, where
$L_{M}^{(1)}=-\sqrt{2} \sum_{\alpha} A_{M}^{1}(\alpha,\alpha)$ and
$I_{M}^{(1)}=[a^{\dag}_{j}a_{j}]^{(1)}_{M}$. The second term in
(\ref{tm1}) changing the number of bosons by $\Delta N=2$ comes also
from the symplectic extension. In (\ref{tm1}) $g$ and $g_{FG}$ stand
for the gyromagnetic factors, which we consider as free parameters.

The $B(M1)$ values can be obtained from the reduced matrix elements
of $M1$ operator in the usual way:
\begin{equation}
B(M1;J\rightarrow J')=\frac{1}{2J+1}\mid \langle \quad
\gamma',J'\parallel T^{M1}\parallel \gamma,J \quad\rangle \mid ^{2}.
\label{defM1}
\end{equation}
The labels $\gamma$ and $\gamma'$ denote the quantum numbers of the
basis states in chain (\ref{chain}).

For the calculation of the matrix element of first term in
(\ref{tm1}) we note that the $J_{m}^{(1)}$ operator is a generator
of $Spin^{BF}(3)$ algebra. Hence, the Wigner-Eckart theorem can be
applied at the $Spin^{BF}(3)$ level. Using the latter, one obtains
for the required reduced matrix element
\begin{equation}
\langle \gamma',J' || J_{m}^{(1)} || \gamma, J \rangle =
\delta_{\gamma,\gamma'} \delta_{J,J'} \sqrt{J(J+1)(2J+1)}.
\label{MEofJ}
\end{equation}

Further, we will calculate the matrix element of
$F_{(0,1)[0]_{2}\quad 00}^{\quad \lbrack 2]_{6}\quad \ 10}$ of the
second term in (\ref{tm1}) which is a generator of $Sp(12,R)$
algebra. The action of the latter concerns only the orbital part of
the basis functions (\ref{Basis}). In general, for calculating the
matrix elements of symplectic generators, we have the advantage of
using the generalized Wigner-Eckart theorem in two steps \cite{TP}.
For the $SU(3)\rightarrow SO(3)$ and $SU(2)\rightarrow U(1)$
reduction we need the standard $SU(2)$ Clebsch-Gordan coefficients
\begin{widetext}
\begin{equation}
\begin{tabular}{l}
$\langle \lbrack N^{\prime }]\, (\lambda ^{\prime },\mu ^{\prime
});K^{\prime }L^{\prime }M^{\prime };T^{\prime }T_{0}^{\prime
}|T_{[\sigma ]_{3}[2t]_{2}\quad tt_{0}}^{\quad \lbrack \chi ]_{6}\,
\quad lm}|[N]\,
(\lambda ,\mu );KLM;TT_{0}\,\rangle $ \\
\\
$=\langle \lbrack N^{\prime }](\lambda ^{\prime },\mu ^{\prime
});K^{\prime }L^{\prime }||T_{[\sigma ]_{3}[2t]_{2}\quad
tt_{0}}^{\quad \lbrack \chi ]_{6}\,\quad lm}||[N](\lambda ,\mu
);KL\rangle C_{LMlm}^{L^{\prime }M^{\prime
}}C_{TT_{0}tt_{0}}^{T^{\prime }T_{0}^{\prime }}.$
\end{tabular}
 \label{ME}
\end{equation}
\end{widetext}
For the calculation of the double-barred reduced
matrix elements in (\ref{ME}) we use the next step:
\begin{widetext}
\begin{equation}
\begin{tabular}{l}
$\langle \lbrack N^{\prime }]\, (\lambda ^{\prime },\mu ^{\prime
});K^{\prime }L^{\prime }||T_{[\sigma ]_{3}[2t]_{2}\quad
tt_{0}}^{\quad \lbrack \chi ]_{6}\quad \quad lm}||[N]\,(\lambda
,\mu );KL\rangle $ \\
\\
$=\langle \lbrack N^{\prime }]|||T_{[\sigma ]_{3}[2t]_{2}}^{\quad
\lbrack \chi ]_{6}}|||[N]\rangle C_{(\lambda ,\mu )[2T]_{2}\quad
\lbrack \sigma ]_{3}[2t]_{2}\quad (\lambda ^{\prime },\mu ^{\prime
})[2T^{\prime }]_{2}}^{[N]_{6\ }\ \ \ \ \ \ \ \ \ \ \ [\chi ]_{6}\ \
\ \ \ \ \ \ \ \ [N^{\prime }]_{6\ }}C_{KL\quad k(l)_{3}\quad
K^{\prime }L^{\prime }}^{(\lambda ,\mu ) \quad [\lambda ]_{3} \quad
(\lambda ^{\prime },\mu ^{\prime })}$,
\end{tabular}
\label{3-barredME}
\end{equation}
\end{widetext}
where $C_{(\lambda ,\mu )[2T]_{2}\quad \lbrack \sigma
]_{3}[2t]_{2}\quad (\lambda ^{\prime },\mu ^{\prime })[2T^{\prime
}]_{2}}^{[N]_{6\ }\ \ \ \ \ \ \ \ \ \ \ [\chi ]_{6}\ \ \ \ \ \ \ \ \
\ [N^{\prime }]_{6\ }}$ and $C_{KL\quad k(l)_{3}\quad K^{\prime
}L^{\prime }}^{(\lambda ,\mu )\quad[\lambda ]_{3}\quad (\lambda
^{\prime },\mu ^{\prime })}$ are $U(6)$ and $SU(3)$ isoscalar
factors (IF's). Obviously the practical value of the application of
the generalized Wigner-Eckart theorem for the calculation of the
matrix elements of the $Sp(12,R)$ generators depends on the
knowledge of the isoscalar factors for the reductions $U(6)\supset
U(3)\otimes U(2)$ and $U(3) \supset O(3)$, respectively. For the
evaluation of the matrix elements (\ref{ME}) of the $Sp(12,R)$
operators in respect to the chain (\ref{uchain}) the  reduced
triple-barred $U(6)$ matrix elements are also required
(\ref{3-barredME}).

Thus, for the calculation of the matrix element
\begin{widetext}
\begin{equation}
\begin{tabular}{l}
$\langle \lbrack N+2],(0,\mu +1);0L+10;00|F_{(0,1)[0]_{2}\quad
00}^{\quad \lbrack 2]_{6}\quad \quad 10}|[N],(0,\mu );0L0;00\rangle $ \\
\\
$=C_{(0,\mu )[0]_{2}\quad (0,1)[0]_{2}\quad (0,\mu
+1)[0]_{2}}^{[N]_{6\ }\ \ \ \ \ \ \ [2]_{6}\ \ \ \ \ \ \ \ [N+2]_{6\
}} C_{\ \ L\ \quad \ 1\ \quad \ \ \ L+1}^{(0 ,\mu )\ \ (0,1)\ \ \
(0,\mu+1)}
C_{L,0\ \ \ \ \ 1,0}^{L+1,0}$ \\
\\
$\times \langle \lbrack N+2] \ ||| \ F_{(0,1)[0]_{2}}^{\quad \lbrack
2]_{6}\quad }||| \ [N]\rangle $
\end{tabular}
\label{RMEF}
\end{equation}
\end{widetext}
we use the standard recoupling technique for two
coupled $U(6)$ tensors \cite{TP}, \cite{Recoupling}:
\begin{equation}
\begin{tabular}{l}
$\langle \lbrack N^{\prime }]|||\ [T^{[\alpha ]_{6}}\times T^{[\beta
]_{6}}]^{\sigma \lbrack \gamma ]_{6}}\ |||[N]\rangle $ \\
\\
$=\underset{c,\rho _{1},\rho _{2}}{\sum }U([N]_{6};[\beta
]_{6};[N^{\prime }]_{6};[\alpha ]_{6}|[N_{c}]_{6}\rho _{2}\rho
_{1};[\gamma ]_{6}\ \sigma )$
\\
\\
$\times \langle \lbrack N^{\prime }]|||\ T^{[\alpha ]_{6}}\
|||[N_{c}]\rangle \langle \lbrack N_{c}]|||T^{[\beta ]_{6}}\ |||[N]\rangle ,$%
\end{tabular}
\label{RCF}
\end{equation}%
where $U(...)$ are the $U(6)$ Racah coefficients in unitary form
\cite{UNRacah}. For the reduced triple-barred matrix element in our
case, which is multiplicity free and hence there is no sum, we have
\begin{equation}
\begin{tabular}{l}
$\langle \lbrack N+2]||| \ F_{(0,1)[0]_{2}}^{\quad \lbrack
2]_{6}\quad }|||[N]\rangle $ \\
\\
$=U([N]_{6};[1]_{6};[N+2]_{6};[1]_{6}|[N+1]_{6};[2]_{6})$ \\
\\
$\times \langle \lbrack N+2]|||\ u^{\dagger \ [1]_{6}}\
|||[N+1]\rangle
\langle \lbrack N+1]|||\ u^{\dagger \ [1]_{6}}\ |||[N]\rangle $ \\
\\
$=\sqrt{(N+1)(N+2)}$
\end{tabular}
\end{equation}
where the corresponding Racah coefficient for maximal coupling
representations is equal to unity \cite{TP},\cite{Recoupling}. For
obtaining this, we used the fact that in the case of vector bosons
which span the fundamental irrep $[1]$ of $u(n)$ algebra the
$u(n)$-reduced matrix element of raising generators has the well
known form \cite{LeRo}
\begin{equation}
\langle \lbrack N+1]|||\ u_{m}^{\dagger }(\alpha )\ |||[N]\rangle
=\sqrt{N+1}. \label{RTMEofu}
\end{equation}
Taking into account the fact that the corresponding $U(6)$ IF
entering in (\ref{RMEF}) for maximal coupling representations is
equal to 1 \cite{pair},\cite{TP}, we obtain
\begin{widetext}
\begin{equation}
\begin{tabular}{l}
$\langle \lbrack N+2],(0,\mu +1);0L+10;00|F_{(0,1)[0]_{2} \
00}^{\quad \lbrack 2]_{6}\quad 10}|[N],(0,\mu );0L0;00\rangle $ \\
\\
$= C_{L,0\ \ \ \ \ 1,0}^{L+1,0} \ \
[\frac{(\mu+L+3)(L+1)}{(\mu+1)(2L+3)}]^{1/2}\sqrt{(N+1)(N+2)}.$ \\
\end{tabular}
\end{equation}
\end{widetext}
The value of the reduced $SU(3)$ Clebsch-Gordan
coefficient (IF) is taken from Ref.\cite{Ver}. Finally, the yrast
condition $N=2(\mu_{0}+L)=N_{0}+2L$ (or $\mu=\mu_{0}+L$) leads to
the following reduced matrix element
\begin{widetext}
\begin{equation}
\begin{tabular}{l}
$\langle \lbrack N+2],(0,\mu +1);0L+1;00||F_{(0,1)[0]_{2} \
00}^{\quad \lbrack 2]_{6} \quad 10}||[N],(0,\mu );0L;00\rangle $ \\
\\
$= [\frac{(N_{0}+4L+6)(L+1)}{(N_{0}+2L+2)(2L+3)}]^{1/2}
\sqrt{(N_{0}+2L+1)(N_{0}+2L+2)},$ \\
\end{tabular}
\label{MEofF}
\end{equation}
\end{widetext}
where in (\ref{MEofF}) the relation $N=2\mu+\lambda$
is taken into account. We see that the expression (\ref{MEofF})
depends on the ground state collective structure $N_{0}$. If
$N_{0}=0$ (hence $N=2L$), the matrix element reduces simply to
\begin{widetext}
\begin{equation}
\begin{tabular}{l}
$\langle \lbrack N+2],(0,\mu +1);0L+1;00||F_{(0,1)[0]_{2} \
00}^{\quad \lbrack 2]_{6}\quad 10}||[N],(0,\mu );0L;00\rangle $ \\
\\
$= \sqrt{(2L+1)(2L+2)}$ \\
\end{tabular}
\end{equation}
\end{widetext}
obtained in \cite{TP}.

\begin{figure}[h]
\centerline{\epsfxsize=3.5in\epsfbox{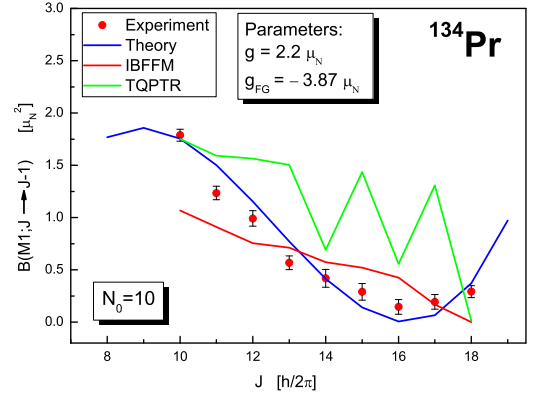}} \caption{(Color
online) Comparison of the theoretical and experimental values for
the in-band $B(M1)$ transition probabilities between the states of
the yrast band for the $^{134}Pr$. The theoretical predictions of
the IBFFM and TQPTR are shown as well.} \label{PrM1}
\end{figure}

For the calculation of the matrix element of $G_{(1,0)[0]_{2}\quad
00}^{\quad \lbrack -2]_{6}\quad  10}$ we use the conjugation
property
\begin{widetext}
\begin{equation}
\begin{tabular}{l}
$\langle \lbrack N-2],(0,\mu -1);0L-1;00||G_{(1,0)[0]_{2} \
00}^{\quad \lbrack -2]_{6}\quad 10}||[N],(0,\mu );0L;00\rangle $ \\
\\
$=(\langle \lbrack N],(0,\mu );0L;00||F_{(0,1)[0]_{2} \ 00}^{\quad
\lbrack 4]_{6}\quad 10}||[N-2],(0,\mu -1);0L-1;00\rangle
)^{\ast }$ \\
\\
$=C_{(0,\mu -1)[0]_{2}\quad (0,1)[0]_{2}\quad (0,\mu
)[0]_{2}}^{[N-2]_{6\ }\ \ \ \ \ \ \ \ \ \ [2]_{6}\ \ \ \ \ \ \
[N]_{6\ }}C_{\ \ L-1\ \quad \ 1\ \quad \ \ \ L}^{(0 ,\mu-1 )\ \
(0,1)\ \ \ (0,\mu)} \sqrt{N(N-1)}$ \\
\\
 $= [\frac{(N_{0}+4L+2)L}{(N_{0}+2L)(2L+1)}]^{1/2}
\sqrt{(N_{0}+2L)(N_{0}+2L-1)}.$ \\
\end{tabular}
\label{MEofG}
\end{equation}
\end{widetext}%
With the help of the above analytic expressions (\ref{MEofJ}),
(\ref{MEofF}) and (\ref{MEofG}) one obtains the corresponding
$B(M1;J \rightarrow J-1)$ values between the states in the yrast
band as attributed to the $SU(3)$ symmetry-adapted basis states of
the model (\ref{Basis}). The numerical values obtained by fitting
the two parameters $g$ and $g_{FG}$ to the experimental data for
$^{134}Pr$ are given in Figure \ref{PrM1}. For comparison, the IBFFM
and TQPTR results (ref. \cite{DC1}) are also shown. From the figure
one can see that while the IVBM and IBFFM results in the $J \approx
13-17$ region are with almost the same level of accuracy, the
general experimental trend is fairly well reproduced in the
framework of the former. The adopted values of effective g-factors
are $g = 2.2 \mu_{N}$ and $g_{FG} = -3.87 \mu_{N}$.

\begin{figure}[h]
\centerline{\epsfxsize=3.5in\epsfbox{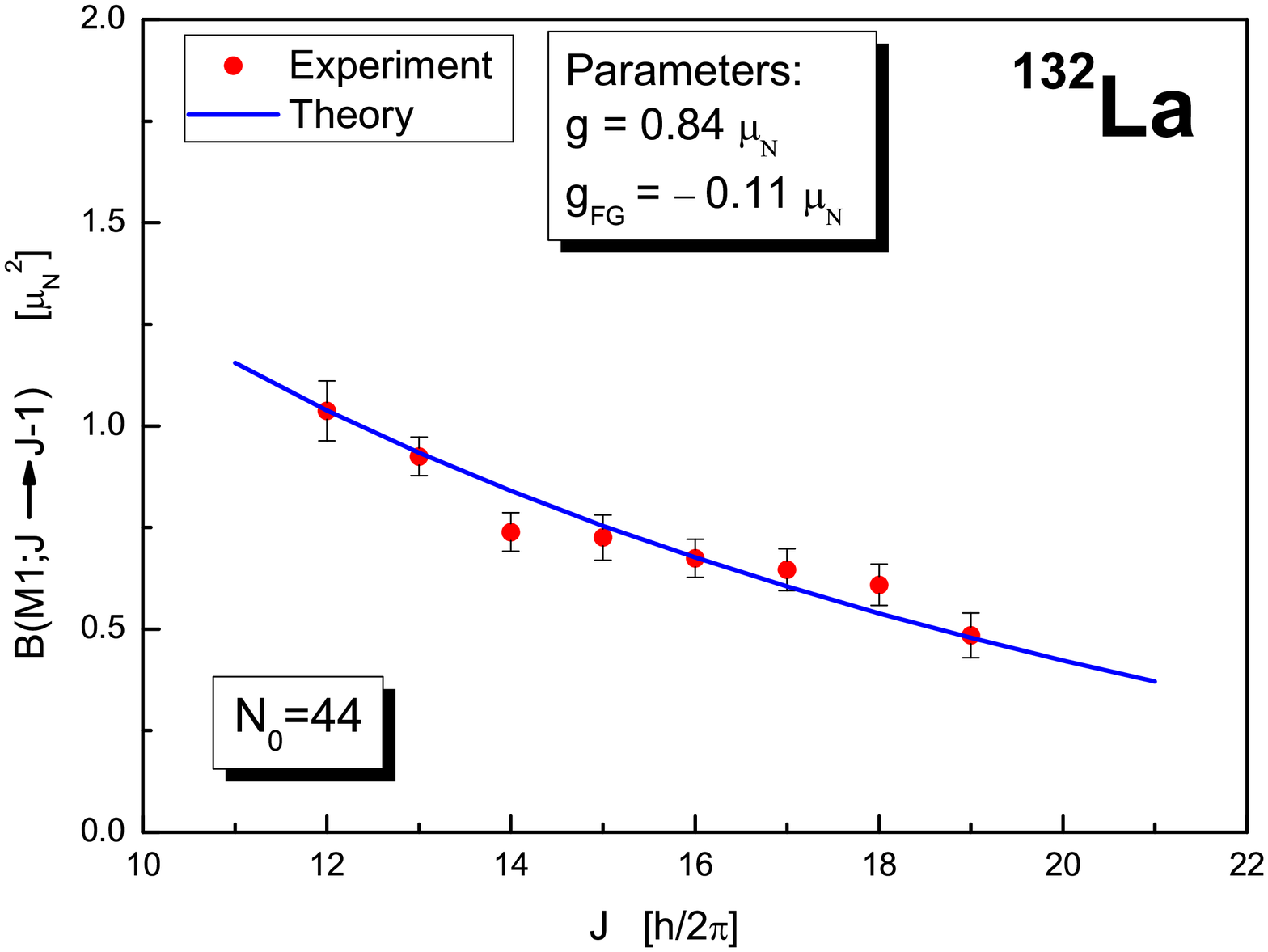}} \caption{(Color
online) Comparison of the theoretical and experimental values for
the in-band $B(M1)$ transition probabilities between the states of
the yrast band for the $^{132}La$.} \label{LaM1}
\end{figure}

In Figure \ref{LaM1} the theoretical predictions for the $^{132}La$
nucleus are compared with experiment \cite{e3}. The adopted values
of effective g-factors are $g = 0.84 \mu_{N}$ and $g_{FG} = -0.11
\mu_{N}$. One can observe a very good description of the
experimental data within the framework of the present approach in
this nucleus as well. We want to point out that the contribution of
the symplectic term in (\ref{defM1}) is crucial for the accurate
reproduction of the experimental $B(M1)$ behavior.

The calculation of the $M1$ and $E2$ transitions in the side band
requires the knowledge of the corresponding $U(6)$ and $SU(3)$
isoscalar factors which are not available analytically for the basis
states attributed to states of the side band. The computer codes
\cite{Dra} for the numerical calculation of the $SU(3)$ IF's can be
used, so the difficulties are focused on the calculation of the
corresponding $U(6)$ isoscalar factors. Hence, the calculation of
the transition probabilities in the side band is a nontrivial task
for a future application of our approach.

\section{Conclusions}

In the present paper, the yrast and yrare states with the $\pi
h_{11/2} \otimes \nu h_{11/2}$ configuration in the doubly odd
nuclei,  $^{126}Pr$,$^{134}Pr$ and $^{132}La$, were investigated in
terms of the orthosymplectic extension of the IVBM. This allows for
the proper reproduction of the energies of these states up to high
angular momenta in both bands.

The basis states of the odd-mass and odd-odd systems are classified
by the dynamical symmetry (\ref{chain}) and the model Hamiltonian is
written in terms of the first and second order invariants of the
groups from the corresponding reduction chain. Hence the problem is
exactly solvable within the framework of the IVBM which, in turn,
yields a simple and straightforward application to real nuclear
systems.

For two of the three isotopes considered, the $B(E2)$ and $B(M1)$
transition probabilities between the states of the yrast band are
calculated and compared with the experimental data. A good overall
agreement of the theoretical predictions with experiment is
obtained. The calculations reveal the important role of the
symplectic term entering in the corresponding transition operator
for the correct reproduction of the behavior of both $B(E2)$ and
$B(M1)$ strengths.

The even-even nuclei are used as a core on which the collective
excitations of the neighboring odd-mass and odd-odd nuclei are build
on. Thus, the spectra of odd-mass and odd-odd nuclei arise as a
result of the coupling of the fermion degrees of freedom to the
boson core. The states of the yrast band of doubly odd nuclei are
build on the ground state band $SU(3)$ multiplets $(0,\mu)$ of the
even-even core, while those of side band are build on the $SU(3)$
multiplets $(\lambda_{0}=fixed,\mu)$ which suggest similar
collective behavior (both sets of $SU(3)$ multiplets are the
stretched states of second type) of the two bands. The only
difference comes from the initial band head structures ($(0,\mu)$
and $(\lambda_{0},\mu)$).  Hence, a purely collective structure of
the states of the yrast and side bands is introduced. Those
assumptions suggest a similar behavior (slope) of the transitions in
the side band which we intend to investigate in future.

The good agreement between the theoretical and the experimental band
structures is a result of the mixing of the basic collective modes
$-$rotational and vibrational ones arising from the yrast
conditions, way back on the level of the even-even cores. This
allows for the correct reproduction of the high spin states of the
collective bands and the correct placement of the different band
heads. The simplifications in our approach comes from the fact that
only one dynamical symmetry is employed, which leads to exact and
simple solutions depending only on the values of the model
parameters. The success of the presented applications is based on
the proper and consistent mapping of the experimentally observed
collective states of the even-even, odd-mass and odd-odd nuclei on
the (ortho)symplectic structures. The latter is much simpler
approach than the mixing of the basis states considered in other
theoretical models.

The presented results on the description of the doublet bands in
odd-odd nuclei confirm the wider applicability of the used
boson-fermion symmetry of IVBM. \\

\section*{Acknowledgments}

This work was supported by the Bulgarian National Foundation for
scientific research under Grant Number $\Phi -1501$. H. G. G.
acknowledges also the support from the European Operational programm
HRD through contract $BG051PO001/07/3.3-02$ with the Bulgarian
Ministry of Education.


\begin{thebibliography}{99}

\bibitem{a1} K. Starosta et al., Phys. Rev. Lett. \textbf{86}, 971
(2001).
\bibitem{a2} A. A. Hecht et al., Phys. Rev. \textbf{C 63}, 051302
(2001).
\bibitem{a3} T. Koike et al., Phys. Rev. \textbf{C 63}, 061304
(2001).
\bibitem{a4} D. J. Hartley et al., Phys. Rev. \textbf{C 64}, 031304
(2001).
\bibitem{a5} R. A. Bark et al., Nucl. Phys. \textbf{A 691}, 577
(2001).
\bibitem{a6} K. Starosta et al., Phys. Rev. \textbf{C 65}, 044328
(2002).

\bibitem{a7} T. Koike et al., Phys. Rev. \textbf{C 67}, 044319
(2003).
\bibitem{a8} G. Rainovski et al., Phys. Rev. \textbf{C 68}, 024318
(2003).

\bibitem{FM} S. Frauendorf, J. Meng, Nucl. Phys. \textbf{A 617}, 131 (1997).

\bibitem{tac1} K. Starosta et. al., Phys. Rev. Lett. \textbf{86}, 971 (2001).

\bibitem{tac2} A.A. Hecht et. al., Phys. Rev. \textbf{C 63}, 051302(R)
(2001).

\bibitem{tac4} V.I. Dimitrov, S. Frauendorf, F. D¨onau, Phys. Rev. Lett.
\textbf{84}, 5732 (2000).

\bibitem{cqpcm2} A.J. Simons et. al., J. Phys. G \textbf{3}1, 541 (2005).

\bibitem{prm1} D.J. Hartleyet et. al., Phys. Rev. \textbf{C 64}, 031304(R) (2001).

\bibitem{prm2} R.A. Bark et. al., Nucl. Phys. \textbf{A 691}, 577 (2001).

\bibitem{prm3} T. Koike, K. Starosta, I. Hamamoto, Phys. Rev. Lett. \textbf{93}, 172502
(2004).

\bibitem{tqptrm} I. Ragnarsson and P. Semmes, Hyperfine Interact. \textbf{43},
423 (19988).

\bibitem{pr134} S. Brant, D. Vretenar, and A. Ventura, Phys. Rev. \textbf{C 69}, 017304 (2004).

\bibitem{e2} E. Grodner, J. Srebrny, Ch. Droste, T. Morek, A. Paster-
nak, J. Kownacki, Int. J. Mod. Phys. E \textbf{13}, 243 (2004).

\bibitem{e3} J. Srebrny et al.,
Acta Phys. Pol. B \textbf{36}, 1063 (2005).

\bibitem{e4} D. Tonev et. al., Phys. Rev. Lett. \textbf{96}, 052501
(2006).

\bibitem{e5} C.M. Petrache, G.B. Hagemann, I. Hamamoto,
K. Starosta, Phys. Rev. Lett. \textbf{96}, 112502 (2006).

\bibitem{ptsm1} K. Higashiyama, N. Yoshinaga, Prog. Theor. Phys. \textbf{113}, 1139 (2005).

\bibitem{ptsm2} K. Higashiyama, N. Yoshinaga, K. Tanabe, Phys. Rev. \textbf{C
72}, 024315 (2005).

\bibitem{ptsm3} N. Yoshinaga, K. Higashiyama, J. Phys. G \textbf{31}, S1455
(2005).

\bibitem{qcm1} N. Yoshinaga, K. Higashiyama, Eur. Phys. J. A \textbf{30}, 343
(2006); \textbf{31}, 395 (2007).

\bibitem{qcm2} K. Higashiyama and N. Yoshinaga, Eur. Phys. J. A \textbf{33},
355 (2007).

\bibitem{IBFFMa} V. Paar, \textit{Capture Gamma-Ray Spectroscopy and Related Topics}, edited by
S. Raman, AIP Conf. Proc. No. 125 (AIP, New York, 1985), p. 70; S.
Brant, V. Paar, and D. Vretenar, Z. Phys. \textbf{A319}, 355 (1984);
V. Paar, D. K. Sunko, and D. Vretenar, Z. Phys. \textbf{A327}, 291
(1987).

\bibitem{IBFFMb} S. Brant and V. Paar, Z. Phys. \textbf{A329}, 151
(1988).

\bibitem{DC1} D. Tonev et al., Phys. Rev. \textbf{C 76}, 044313 (2007).

\bibitem{DC2} S. Brant, D. Tonev, G. de Angelis, and A. Ventura,
Phys. Rev. \textbf{C 78}, 034301 (2008).

\bibitem{pr126} C. M. Petrache et al., Phys. Rev. \textbf{C 64}, 044303 (2001).

\bibitem{GGG} H. Ganev, V. P. Garistov, and A. I. Georgieva,
Phys. Rev. C \textbf{69}, 014305 (2004).

\bibitem{OSE} H. G. Ganev, J. Phys. G: Nucl.
Phys. \textbf{35}, 125101 (2008).

\bibitem{OON} H. G. Ganev and A. I. Georgieva, Proc. of XXVIIth
International Wokrshop on Nuclear Theory, Rila Mountains 2008.

\bibitem{IVBM} A. Georgieva, P. Raychev, and R. Roussev, J. Phys. G: Nucl.
Phys. \textbf{8}, 1377 (1982).

\bibitem{Q1} C. Quesne, I. Math. Phys. \textbf{14}, 366 (1973).

\bibitem{str} D. J. Rowe, Rep. Prog. Phys. \textbf{48}, 1419
(1985).

\bibitem{GGD} H. G. Ganev, V. P. Garistov, A. I. Georgieva, and J. P. Draayer, Phys. Rev.
\textbf{C 70}, 054317 (2004).

\bibitem{exp} Evaluated Nuclear Structure Data File (ENSDF),
http://ie.lbl.gov/databases/ensdfserve.html; Level Retrieval
Parameters, http://iaeand.iaea.or.at/nudat/levform.html.

\bibitem{WJR} L. Wilets and M. Jean, Phys. Rev.
\textbf{102}, 788 (1956).

\bibitem{Pr130132} C.M. Petrache, S. Brant, D. Bazzacco, G. Falconi, E. Farnea, S.
Lunardi, V. Paar, Zs. Podolya´k, R. Ventureli, and D. Vretenar,
Nucl. Phys. \textbf{A635}, 361 (1998).

\bibitem{pair} B. G. Wybourne,
\textit{Classical Groups for Physicists} (Wiley, New York, 1974).

\bibitem{Elliott} J. P. Elliott, Proc. R. Soc. \textbf{A245}, 128, 562 (1958).

\bibitem{TP} H. G. Ganev and A. I. Georgieva, Phys. Rev.
\textbf{C 76}, 054322 (2007).

\bibitem{ng} H. G. Ganev, to be published.

\bibitem{Recoupling} B. G. Wybourne, Classical Groups for Physicists,
(Wiley, New York, 1974); C. Quesne, J. Phys. A: Math. Gen.
\textbf{23}, 847 (1990); \textbf{24}, 2697 (1991).

\bibitem{UNRacah} K. T. Hecht, R. Le Blanc and D. J. Rowe, J. Phys. A: Math.
Gen. \textbf{20}, 2241 (1987).

\bibitem{LeRo} R. Le Blanc and D. J. Rowe, J. Phys. A: Math.
Gen. \textbf{20}, L681 (1987).

\bibitem{Ver} D. Vergados, Nucl. Phys. \textbf{A111}, 681 (1968).

\bibitem{Dra} J. P. Draayer and Y. Akiyama, J. Math. Phys.
\textbf{14}, 1904 (1973); Y. Akiyama and J. P. Draayer, Comput.
Phys. Commun. \textbf{5}, 405 (1973); C. Bahri, J. P. Draayer,
Comput. Phys. Commun. \textbf{83}, 59 (1994); C. Bahri, D. J. Rowe,
J. P. Draayer, Comput. Phys. Commun. \textbf{159}, 121 (2004).

\end{thebibliography}
\end{document}